\documentclass[conference]{IEEEtran}
\IEEEoverridecommandlockouts
\usepackage{algcompatible}
\usepackage{amssymb}
\usepackage{amsfonts}
\usepackage{amsmath}
\usepackage{amsthm}
\usepackage{blkarray}
\usepackage{blindtext}
\usepackage{graphicx}
\usepackage{booktabs}
\usepackage{dsfont}
\usepackage{balance}
\usepackage{color}
\usepackage{cite}
\usepackage{enumerate}
\usepackage{footnote}
\usepackage{graphicx}
\usepackage{verbatim}
\usepackage{stfloats}
\usepackage{mdframed}
\usepackage{multirow}
\usepackage{subfigure}
\usepackage{tablefootnote}
\usepackage{tabularx} 
\usepackage{textcomp}
\usepackage{float}
\usepackage{xr}
\usepackage{xcolor}
\usepackage[final]{hyperref} 
\usepackage[lined, boxed, linesnumbered, ruled]{algorithm2e}
\newtheorem{theorem}{Theorem}[section]

\newtheorem{definition}{Definition}
\newtheorem{lemma}{Lemma}
\newtheorem{proposition}{Proposition}[section]

\usepackage[utf8]{inputenc}

\usepackage[ letterpaper, left=0.625in, right=0.625in, top=0.75in, bottom=1in ]{geometry}
\title{Theoretical Delay Analysis of Network Coding Enabled M-to-N Broadcasting in Ad-Hoc Networks}

\author{
  \IEEEauthorblockN{Zhaohong Lu\IEEEauthorrefmark{1}, Qingyu Liu\IEEEauthorrefmark{2},Haibo Zeng\IEEEauthorrefmark{1}}
  \IEEEauthorblockA{\IEEEauthorrefmark{1} Dept. of Electrical and Computer Engineering, Virginia Tech, Blacksburg, VA 24060, USA\\
                    \IEEEauthorrefmark{2}Shenzhen Graduate School, School of Electric and Computer Engineering, Peking University, Shenzhen 518055, China\\
                    Emails: \{zhaohonglu,hbzeng\}@vt.edu, qy.liu@pku.edu.cn
                    }
}

\begin{document}

\maketitle

\begin{abstract}
In this paper, we investigate the delay performance of ad hoc broadcast networks in which random linear network coding is used to support M-to-N information dissemination. Multiple source nodes may inject coded packets for the same message, while multiple destination nodes decode the message after collecting a sufficient number of innovative degrees of freedom. Because broadcast forwarding does not rely on predetermined end-to-end routes, the packet-level relay sequence is generally random and depends on propagation delay, queueing delay, and prior reception history. We first prove that, when queueing, contention, processing, and retransmission delays are ignored, the first-reception time under fastest-only broadcast is exactly equal to the multi-source shortest-path distance, which provides a lower bound on propagation delay. We then develop an equivalent fixed-route queueing approximation to estimate relay-node load, stability, and end-to-end delay under M/D/1 and G/D/1 models. To account for route diversity under congestion, we further introduce a congestion-cost path approximation that penalizes heavily loaded relays. Simulation results on random and structured topologies show that the proposed framework captures the main delay trends under different traffic intensities and network densities while remaining simple for tractable analytical evaluation.
\end{abstract}

\section{Introduction}

Many applications, such as vehicle-to-everything (V2X) communication and wireless sensor networks, require timely message dissemination over dynamic multi-hop networks. Mobility, limited transmission range, channel variation, and temporary node failures make accurate routing-table maintenance costly and unreliable. Safety-related messages are strongly delay-sensitive and may lose their value if delivered too late \cite{SHAHWANI2022100420}. The problem becomes more severe under heavy traffic, since channel congestion can cause message drops, increased delay, and degraded quality of service \cite{PARANJOTHI2020102277}. Moreover, practical V2X messages often need to reach vehicles beyond one-hop range and therefore rely on multi-hop broadcasting \cite{9775109}. Opportunistic routing, coding, and broadcast forwarding can maintain dissemination without stable end-to-end routes, but they also introduce redundancy and uncertainty that complicate delay estimation.

Opportunistic routing relaxes fixed-path dependence by selecting the next forwarder after a packet is received by one or more candidate relays \cite{10.1145/972374.972387,10.1145/1282380.1282400}. Anypath routing extends this principle by allowing each node to broadcast to a forwarding set, with the realized path determined by the successful receivers \cite{5061904}. Coding-based broadcasting provides a complementary direction. For example, rateless coding exploits broadcast reception so that relays transmit only fractions of encoded data, reducing redundant transmissions while maintaining reliable one-to-all dissemination \cite{10.1145/1280940.1280959}.

Another major approach combines random linear network coding (RLNC) with broadcast forwarding. RLNC is well suited to route-free dissemination because each innovative packet contributes one degree of freedom, regardless of its source or relay sequence. Foundational studies established coding gains over routing and showed that distributed RLNC can achieve multicast capacity with high probability \cite{850663,1228459,1705002}. In wireless ad-hoc networks, coding reduces transmissions for all-to-all broadcast, deterministic broadcast, and reliable multicast \cite{4146698,4215785,4753100}. Practical protocol studies further considered realistic MAC behavior, unreliable links, scheduling, deadlock, reliability, and retransmission overhead \cite{5403547,5426175}. More recent work extended this direction to multi-source, many-to-all, mesh, and UAV networks, reporting improvements in delivery, latency, and transmission efficiency \cite{6363697,10.1145/3274783.3274849,9201370}.

These studies demonstrate the benefits of RLNC-based broadcasting, since a receiver needs innovative degrees of freedom rather than a specific packet from a specific source along a predetermined route. However, most existing work emphasizes reliability, reachability, transmission reduction, or protocol operation. When relay nodes have finite service capacity, broadcast dissemination can still create queues, bottlenecks, and load-dependent delays. Analyzing this delay in an M-to-N route-free broadcast network therefore remains challenging.

Delay performance has also been studied in RLNC-based wireless broadcast systems. Existing work has analyzed throughput--delay scaling with coding-window size \cite{6553237}, decoding delay under deadline constraints \cite{7858628}, delay--complexity tradeoffs in wireless broadcast \cite{9112345}, completion delay in full-duplex relay networks \cite{9944003}, and delay minimization with network coding and feedback over erasure broadcast channels \cite{5205612}. These studies provide important insights into how coding parameters, erasure channels, deadlines, feedback, and relay operations affect the delay of RLNC-based transmission. However, they do not directly characterize delay in general multi-hop route-free RLNC broadcast networks.

There has also been substantial progress in end-to-end delay analysis for multi-hop and wireless networks. Neely and Modiano studied capacity-delay tradeoffs in mobile ad hoc networks and showed how redundant packet transfers can reduce delay at the cost of additional congestion \cite{1435642}. Queuing-network models have also been used to analyze average end-to-end delay in random-access multihop wireless ad hoc networks \cite{10.1145/1143549.1143704}. Gupta and Shroff derived delay lower bounds for multi-hop wireless networks with fixed routes \cite{5688207}, and delay modeling has also been studied for traffic flows traversing virtual network function chains in 5G networks \cite{8408468}. These works  provide important analytical foundations for network delay characterization. Nevertheless, their fixed-route assumptions do not directly apply to RLNC-based broadcast, where the packet-level relay sequence is not known in advance and may depend jointly on propagation delay, relay queueing delay, and reception history. 

Motivated by this gap, this paper develops a tractable delay-analysis framework for RLNC-enabled M-to-N broadcasting with finite FIFO relay service. We first prove that, when queueing and other load-dependent effects are ignored, fastest-only broadcast yields the multi-source shortest-path first-reception time, providing a rigorous lower bound. We then develop a bottleneck-reduction shortest-path approximation that estimates broadcast-induced relay load and queueing delay through M/D/1 and G/D/1 models. Finally, a congestion-cost detour approximation explains cases in which dissemination follows longer but less congested relay sequences. These components provide a systematic way to estimate delay, identify relay bottlenecks, and characterize load-dependent route changes.

The main contributions of this paper are summarized as follows:
\begin{itemize}
    \item We formulate an M-to-N RLNC broadcast model with multiple sources, multiple destinations, finite relay-service capacity, and innovation-based forwarding. We also prove that queue-free fastest-only broadcast achieves the multi-source shortest-path first-reception time, giving a rigorous lower bound on packet reception delay.

    \item We develop a bottleneck-reduction queueing approximation for route-free broadcast. It maps broadcast-induced relay load to reduced M/D/1 and G/D/1 queues, thereby connecting source injection rates, relay service constraints, and end-to-end first-arrival delay.

    \item We show that network coding removes packet-identity dependence but not relay-service constraints, introduce a congestion-cost detour approximation, and validate the framework on random and structured topologies. The results show that the main estimate captures load-dependent delay trends, while the detour approximation explains route changes around congested relays.
\end{itemize}

The remainder of the paper is organized as follows. Sections~\ref{sec:sys} and~\ref{sec:broadcast} present the model and protocol, Section~\ref{sec:delay} develops the delay analysis, Section~\ref{sec:num} reports the simulations, and Section~\ref{sec:con} concludes the paper.

\section{System Model and Preliminaries}
\label{sec:sys}

\subsection{Network Model}

Consider a time-slotted multi-hop wireless network $\mathbf{G}=(\mathbf{V},\mathbf{E})$, where $\mathbf{V}$ is the set of nodes and $\mathbf{E}$ is the set of links. For each node $v\in\mathbf{V}$, let
\begin{equation}
    \mathcal{N}^{+}(v)=\{u:(v,u)\in\mathbf{E}\}
\end{equation}
denote the set of outgoing neighbors of $v$.

Each link $e=(v,u)\in\mathbf{E}$ has a propagation delay $\tau_e$. When node $v$ broadcasts a packet in slot $t$, each node $u\in\mathcal{N}^{+}(v)$ receives a copy in slot $t+\tau_{(v,u)}$. Each link is supposed to be collision-free and noiseless, which means MAC contention, interference, random loss, or retransmission are not explicitly modeled.

Let $\mathcal{M}$ denote the set of messages. For each message $m\in\mathcal{M}$, let $\mathcal{S}_m\subseteq\mathbf{V}$ be the set of source nodes that generate packets for $m$, and let $\mathcal{D}_m\subseteq\mathbf{V}$ be the set of destination nodes that require $m$. We assume
\begin{equation}
    \mathcal{S}_m\cap\mathcal{D}_m=\emptyset.
\end{equation}

All of the messages are distributed using the broadcasting method, where each node can receive and re-broadcast all the packets sent to it. Each node has a single outgoing packet server. Thus, a node may receive packets from its neighbors while transmitting, but it can serve at most one outgoing packet at a time. Let $\mu_v$ denote the service rate of node $v$. Unless otherwise specified, we assume a unit service rate,
\begin{equation}
    \mu_v=1,\quad \forall v\in\mathbf{V}.
\end{equation}

Because the network uses broadcast forwarding, packet-level routes are not fixed in advance. A coded packet may be received through different relay sequences depending on propagation delay, relay queueing delay, and previous reception history. For a source-destination pair $(s,d)$, let $\mathcal{R}_{sd}$ denote the set of feasible directed paths from $s$ to $d$. The shortest-path propagation component used in the delay approximation is defined later in Section~\ref{subsec:shortest_path_component}.

\subsection{Random Linear Network Coding}
Linear network coding is employed during the broadcasting process. The sources encode a message by dividing it into several packets and then generating linear combinations of the packets with random encoding coefficients. To be specific, for each message $m$, the original information is divided into $K_m$ source packets over a finite field $\mathbb{F}_q$:
\begin{equation}
    x_{m,1},x_{m,2},\ldots,x_{m,K_m}.
\end{equation}
A coded packet for message $m$ is a random linear combination
\begin{equation}
    y
    =
    \sum_{k=1}^{K_m}\alpha_k x_{m,k},
\end{equation}
where $\alpha_k\in\mathbb{F}_q$ is a random coding coefficient. The packet is represented by
\begin{equation}
    p=(m,\boldsymbol{\alpha},y),
\end{equation}
where
\begin{equation}
    \boldsymbol{\alpha}
    =
    (\alpha_1,\alpha_2,\ldots,\alpha_{K_m})
    \in\mathbb{F}_q^{K_m}
\end{equation}
is the coding vector.

For each node $v$ and message $m$, let $\mathcal{B}_{v,m}(t)$ denote the set of coding vectors stored by node $v$ by time $t$. The rank accumulated by node $v$ for message $m$ is
\begin{equation}
    r_{v,m}(t)
    =
    \operatorname{rank}\left(\mathcal{B}_{v,m}(t)\right).
\end{equation}
A received coded packet $p=(m,\boldsymbol{\alpha},y)$ is innovative for node $v$ if
\begin{equation}
    \operatorname{rank}\left(
    \mathcal{B}_{v,m}(t)\cup\{\boldsymbol{\alpha}\}
    \right)
    >
    \operatorname{rank}\left(
    \mathcal{B}_{v,m}(t)
    \right).
\end{equation}
Only innovative packets are accepted. Non-innovative packets are discarded.

A destination $d\in\mathcal{D}_m$ can decode message $m$ once it has collected $K_m$ linearly independent coded packets, i.e.,
\begin{equation}
    r_{d,m}(t)\geq K_m.
\end{equation}
Thus, a destination does not need to receive a specific packet from a specific source. It only needs to collect enough innovative degrees of freedom for the message. We assume that packets generated by different sources are always statistic independent.

\subsection{Packet Generation}

Packet generation is modeled in discrete time. Let $A_s^m(t)$ denote whether source node $s\in\mathcal{S}_m$ generates one coded packet for message $m$ in slot $t$:
\begin{equation}
    A_s^m(t)=
    \begin{cases}
        1, & \text{if source } s \text{ generates a coded packet for } m \text{ in slot } t,\\
        0, & \text{otherwise}.
    \end{cases}
\end{equation}
The long-term generation rate of source $s$ for message $m$ is
\begin{equation}
    \lambda_s^m
    =
    \lim_{T\to\infty}
    \frac{1}{T}
    \sum_{t=0}^{T-1}
    \mathbb{E}\left[A_s^m(t)\right].
\end{equation}
The aggregate coded-packet injection rate for message $m$ is
\begin{equation}
    \lambda_m
    =
    \sum_{s\in\mathcal{S}_m}\lambda_s^m.
\end{equation}
This rate represents the rate at which coded packets that may become innovative degrees of freedom at receiving nodes, are injected into the network by the source set $\mathcal{S}_m$.

\section{RLNC Broadcast Protocol}
\label{sec:broadcast}

This section defines the broadcast forwarding protocol used in the delay analysis. The protocol is route-free: nodes do not maintain predetermined end-to-end routes. Instead, each node accepts only innovative coded packets, stores their coding vectors, and forwards accepted packets through its local relay queue. The purpose of the protocol definition is to make the interaction among source generation, relay queueing, broadcast service, duplicate suppression, innovation checking, and decoding explicit.

For each node $v$, let $Q_v(t)$ be its FIFO relay queue at slot $t$. For each message $m$, the rank state $r_{v,m}(t)$ is determined by the coding-vector set $\mathcal{B}_{v,m}(t)$ defined in Section~\ref{sec:sys}. In the protocol, each coded packet is assigned a unique identifier and is written as
\begin{equation}
    p=(\iota_p,m,\boldsymbol{\alpha},y),
\end{equation}
where $\iota_p$ is the packet identifier. To avoid repeated forwarding of the same coded packet, each node maintains a transmitted-packet record $\mathcal{T}_v$, which stores packet identifiers. If packet $p$ is received by node $v$ from node $u$, we denote the immediate previous-hop node by $\mathrm{prev}_v(p)=u$. This record is used only to suppress immediate reverse forwarding of the same packet.

\begin{algorithm}[!t]
\footnotesize
\DontPrintSemicolon
\SetCommentSty{footnotesize}

\textbf{Input:} $\mathbf{G}=(\mathbf{V},\mathbf{E})$, $\mathcal{M}$,
$\{\mathcal{S}_m,\mathcal{D}_m,K_m\}$, $\{\tau_e\}$, $\{\mu_v\}$,
and $T_{\max}$.\\
\textbf{State:} relay queues $Q_v$, buffers $\mathcal{B}_{v,m}$,
transmission records $\mathcal{T}_v$, previous-hop records
$\mathrm{prev}_v(p)$, and reception-event set $\mathcal{A}_{\mathrm{rx}}$.\\

Initialize $Q_v\!\leftarrow\!\emptyset$, $\mathcal{T}_v\!\leftarrow\!\emptyset$,
and $\mathcal{B}_{v,m}\!\leftarrow\!\emptyset$ for all $v,m$.\\
For each $m$ and $s\in\mathcal{S}_m$, initialize $\mathcal{B}_{s,m}$ as a
full-rank basis of $\mathbb{F}_q^{K_m}$.\\
Set $T^{\mathrm{dec}}_{m,d}\!\leftarrow\!\infty$ for all
$d\in\mathcal{D}_m$, and $\mathcal{A}_{\mathrm{rx}}\!\leftarrow\!\emptyset$.\\

\For{$t=0,1,\ldots,T_{\max}$}{
    Initialize temporary updates $\Delta Q_v\leftarrow\emptyset$ and
    $\Delta\mathcal{B}_{v,m}\leftarrow\emptyset$ for all $v,m$.\\

    \tcp{Serve queued packets}
    \For{each node $v\in\mathbf{V}$}{
        \If{$Q_v\neq\emptyset$ and the server of $v$ is available}{
            Pop $p=(\iota_p,m,\boldsymbol{\alpha},y)$ from $Q_v$ and add
            $\iota_p$ to $\mathcal{T}_v$.\\
            \For{each $u\in\mathcal{N}^{+}(v)$ with
            $u\neq\mathrm{prev}_v(p)$}{
                Add $(p,v,u,t+\tau_{(v,u)})$ to
                $\mathcal{A}_{\mathrm{rx}}$.\\
            }
        }
    }

    \tcp{Process packet receptions}
    \For{each $(p,v,u,t_{\mathrm{rx}})\in\mathcal{A}_{\mathrm{rx}}$
    with $t_{\mathrm{rx}}=t$}{
        Remove the event from $\mathcal{A}_{\mathrm{rx}}$ and let
        $p=(\iota_p,m,\boldsymbol{\alpha},y)$.\\
        \If{$\boldsymbol{\alpha}$ is innovative w.r.t.
        $\mathcal{B}_{u,m}\cup\Delta\mathcal{B}_{u,m}$}{
            Add $\boldsymbol{\alpha}$ to $\Delta\mathcal{B}_{u,m}$ and set
            $\mathrm{prev}_u(p)\leftarrow v$.\\
            \If{$\iota_p\notin\mathcal{T}_u$}{
                Append $p$ to $\Delta Q_u$.\\
            }
            \If{$u\in\mathcal{D}_m$,
            $\operatorname{rank}(\mathcal{B}_{u,m}\cup
            \Delta\mathcal{B}_{u,m})\ge K_m$, and
            $T^{\mathrm{dec}}_{m,u}=\infty$}{
                Set $T^{\mathrm{dec}}_{m,u}\leftarrow t$.\\
            }
        }
    }

    \tcp{Generate source packets}
    \For{each $m\in\mathcal{M}$ and $s\in\mathcal{S}_m$}{
        Generate $p=(\iota_p,m,\boldsymbol{\alpha},y)$ according to
        $A_s^m(t)$.\\
        \If{$p$ is generated}{
            Set $\mathrm{prev}_s(p)\leftarrow\emptyset$ and append $p$ to
            $\Delta Q_s$.\\
        }
    }

    \tcp{Apply slot updates}
    \For{each $v\in\mathbf{V}$ and $m\in\mathcal{M}$}{
        Append $\Delta Q_v$ to $Q_v$ in FIFO order, and set
        $\mathcal{B}_{v,m}\leftarrow
        \mathcal{B}_{v,m}\cup\Delta\mathcal{B}_{v,m}$.\\
    }

    \If{$T^{\mathrm{dec}}_{m,d}<\infty$ for all $m$ and
    $d\in\mathcal{D}_m$}{
        Stop.\\
    }
}
\caption{\small Discrete-event RLNC broadcast with FIFO queues and innovation-based forwarding.}
\label{alg:rlnc_broadcast_protocol}
\end{algorithm}

A source node $s\in\mathcal{S}_m$ is regarded as having access to the full $K_m$ source degrees of freedom of message $m$. Therefore, source-generated coded packets are inserted into the source queue without an innovation check at the source. A destination is not removed from the relay process after decoding. Once $d\in\mathcal{D}_m$ reaches rank $K_m$, its decoding time is recorded, but the node may still forward innovative packets for the same message to help other destinations. This convention keeps the broadcast process independent of the order in which different destinations decode.

Algorithm~\ref{alg:rlnc_broadcast_protocol} defines the forwarding process in detail by fixing the slot order, queue discipline, duplicate-forwarding record, innovation check, and decoding-time update. A broadcast consumes one service opportunity at the transmitter and delivers one copy to each outgoing neighbor after the corresponding propagation delay. Non-innovative packets are discarded before entering relay queues, while innovative packets accepted during slot $t$ become eligible for forwarding from slot $t+1$. This convention avoids same-slot forwarding ambiguity and makes the event order well defined.

\subsection{Propagation-Only Lower Bound}

The following proposition analyzes a propagation-only version of Algorithm~\ref{alg:rlnc_broadcast_protocol}. In this idealized case, relay queues are ignored, every node can transmit immediately after first reception, and no service, processing, contention, loss, or retransmission delay is incurred.

\begin{proposition}[Propagation-only lower bound]
\label{prop:propagation_lower_bound}
Consider a static graph $\mathbf{G}=(\mathbf{V},\mathbf{E})$ with link propagation delays $\tau_e$. Under the propagation-only model defined above, the first-reception time $T_v^{\mathrm{prop}}$ of any node $v\in\mathbf{V}$ satisfies
\begin{equation}
    T_v^{\mathrm{prop}}
    =
    \min_{s\in\mathcal{S}_m}
    \operatorname{dist}_{\mathbf{G}}(s,v),
\end{equation}
where $\operatorname{dist}_{\mathbf{G}}(s,v)$ is the shortest-path propagation delay from $s$ to $v$.
\end{proposition}

\begin{proof}
Construct an auxiliary graph $\mathbf{G}'=(\mathbf{V}\cup\{s_0\},\mathbf{E}')$ by adding a virtual source node $s_0$ and zero-delay edges $(s_0,s)$ for all $s\in\mathcal{S}_m$. All original edges keep their propagation delays $\tau_e\geq0$. Let $d(v)$ be the shortest-path distance from $s_0$ to $v$ in $\mathbf{G}'$. By construction,
\begin{equation}
    d(v)
    =
    \min_{s\in\mathcal{S}_m}
    \operatorname{dist}_{\mathbf{G}}(s,v).
\end{equation}

First, no packet can reach $v$ earlier than the accumulated propagation delay of some path from a source in $\mathcal{S}_m$ to $v$. Hence,
\begin{equation}
    T_v^{\mathrm{prop}}\geq d(v).
\end{equation}
Second, consider a shortest path from $s_0$ to $v$ in $\mathbf{G}'$. Since every intermediate node rebroadcasts immediately after its first reception, the packet reaches each node on this path exactly after the accumulated propagation delay along the path. Therefore,
\begin{equation}
    T_v^{\mathrm{prop}}\leq d(v).
\end{equation}
Combining the two inequalities gives $T_v^{\mathrm{prop}}=d(v)$.
\end{proof}

\section{Approximation Analysis of Queueing Delay}
\label{sec:delay}

Proposition~\ref{prop:propagation_lower_bound} gives a propagation-only lower bound on packet delay. In the actual RLNC broadcast protocol, however, a coded packet must also wait for relay service. The relay sequence of a packet is not fixed in advance and may depend on propagation delay, relay backlog, previous receptions, duplicate suppression, and rank evolution. Therefore, an exact delay analysis is highly nontrivial.

This section develops a bottleneck-reduction shortest-path approximation for the queueing delay. The approximation has three components. First, we use a node-service cut argument to show that RLNC removes packet-identity dependence but does not remove service-capacity constraints. Second, we approximate the first-reception propagation component by propagation-only shortest paths. Second, inspired by the bottleneck-reduction method for multi-hop wireless networks \cite{5688207}, we reduce the queueing effect at each relay bottleneck to a single-server G/D/1 queue driven by aggregate exogenous coded-packet arrivals \cite{Kingman_1961}. A congestion-cost detour approximation is also defined as a theoretical extension, which provides a more precise estimation for some special cases.

\subsection{Stability and Load-Induced Bottlenecks}
\label{subsec:stability_bottleneck}

We use the standard notion of mean-rate stability. A node queue is mean-rate stable if its expected backlog grows sublinearly with time \cite{1941130}.

\begin{definition}[Mean-rate stability]
The broadcast queueing network is stable if
\begin{equation}
    \lim_{t\to\infty}
    \frac{\mathbb{E}[Q_v(t)]}{t}
    =
    0,
    \quad \forall v\in\mathbf{V}.
\end{equation}
\end{definition}

Let $\Lambda_v$ denote the effective coded-packet arrival rate assigned to node $v$ in a queueing approximation. The corresponding traffic intensity is
\begin{equation}
    \rho_v
    =
    \frac{\Lambda_v}{\mu_v}.
\end{equation}
A node is called a load-induced bottleneck when $\rho_v$ is close to one. If $\rho_v\geq 1$ under a given approximation, the corresponding single-server queue is unstable and the estimated waiting time becomes unbounded. This instability is a property of the selected approximation. It does not necessarily imply that the original route-free broadcast process is unstable, because the actual broadcast process may exploit relay sequences not represented by a fixed shortest-path abstraction.

\subsection{Node-Service Cut Constraint}
\label{subsec:cut_constraint}

Network coding changes the role of packet identity. A destination does not need a particular packet generated by a particular source; it only needs enough innovative degrees of freedom. However, each innovative degree of freedom still has to be transported by relay transmissions. Therefore, RLNC does not remove service-capacity constraints.

For message $m$ and destination $d\in\mathcal{D}_m$, let
\begin{equation}
    N_{m,d}(t)
    =
    r_{d,m}(t)
\end{equation}
denote the number of innovative degrees of freedom accumulated by destination $d$ by time $t$.

Consider a cut $\Omega\subset\mathbf{V}$ such that $\mathcal{S}_m\subseteq\Omega$ and $d\notin\Omega$. Under the node-server broadcast model, the boundary transmitter set of the cut is
\begin{equation}
    \partial^+\Omega
    =
    \left\{
    v\in\Omega:
    \exists u\notin\Omega \text{ such that } (v,u)\in\mathbf{E}
    \right\}.
\end{equation}
The corresponding node-service cut capacity is
\begin{equation}
    C_{\mathrm{node}}(\Omega)
    =
    \sum_{v\in\partial^+\Omega}\mu_v.
\end{equation}
This capacity counts innovative degrees of freedom, not raw reception events. A single broadcast transmission crossing the cut may be received by multiple nodes outside the cut, but it carries only one global coding vector and can therefore contribute at most one innovative degree of freedom across the cut.

\begin{lemma}[Node-service cut constraint]
\label{lem:node_service_cut}
For any message $m$ and destination $d\in\mathcal{D}_m$,
\begin{equation}
    \limsup_{t\to\infty}
    \frac{\mathbb{E}[N_{m,d}(t)]}{t}
    \leq
    \min_{\Omega:\mathcal{S}_m\subseteq\Omega,\ d\notin\Omega}
    C_{\mathrm{node}}(\Omega).
\end{equation}
\end{lemma}

\begin{proof}
For an innovative coded packet to increase the rank at destination $d$, the corresponding degree of freedom must cross every cut separating $\mathcal{S}_m$ from $d$. Across a cut $\Omega$, only nodes in $\partial^+\Omega$ can transmit coded packets from $\Omega$ to $\mathbf{V}\setminus\Omega$. Since node $v$ can serve at most $\mu_v$ outgoing packets per slot, the total service rate available across the cut is at most $\sum_{v\in\partial^+\Omega}\mu_v$. Although one broadcast transmission may be received by multiple nodes outside the cut, it represents a single coding vector and cannot create multiple independent degrees of freedom by itself. Taking the minimum over all source-destination cuts gives the stated bound.
\end{proof}

Lemma~\ref{lem:node_service_cut} shows that same-message coded packets should be analyzed at the degree-of-freedom level rather than as fixed packet identities. At the same time, queueing delay cannot be ignored when the injected coded-packet rate approaches the service capacity of relay nodes or cuts.

\subsection{Shortest-Path Propagation Component}
\label{subsec:shortest_path_component}

The actual broadcast protocol does not assign a predetermined route to each coded packet. Nevertheless, the propagation-only lower bound in Proposition~\ref{prop:propagation_lower_bound} shows that the earliest possible first reception is governed by shortest-path propagation distance. Therefore, we use propagation-only shortest paths to approximate the first-reception propagation component of the delay.

For each source $s\in\mathcal{S}_m$ and destination $d\in\mathcal{D}_m$, let
\begin{equation}
    R^{\mathrm{sp}}_{s,m,d}
    =
    \arg\min_{R\in\mathcal{R}_{sd}}
    \tau(R)
\end{equation}
be a propagation-only shortest path, where ties are broken deterministically. For a path
\begin{equation}
    R=(v_0,v_1,\ldots,v_\ell),
\end{equation}
where $v_0=s$ and $v_\ell=d$, define its propagation delay as
\begin{equation}
    \tau(R)
    =
    \sum_{i=0}^{\ell-1}\tau_{(v_i,v_{i+1})},
\end{equation}
and its transmitter set as
\begin{equation}
    \mathcal{V}_{\mathrm{tx}}(R)
    =
    \{v_0,v_1,\ldots,v_{\ell-1}\}.
\end{equation}

The shortest path $R^{\mathrm{sp}}_{s,m,d}$ is used only to represent the measured first-arrival propagation path from source $s$ to destination $d$. It does not imply that Algorithm~\ref{alg:rlnc_broadcast_protocol} constrains packets to this route. The queueing component is modeled separately by the bottleneck-reduction load defined next.

\subsection{Node-Bottleneck Reduction}
\label{subsec:node_bottleneck_reduction}

In a multi-hop queueing network, directly characterizing the arrival process at an intermediate relay is difficult because the intermediate arrival process is itself the departure process of upstream queues. The bottleneck-reduction technique avoids this difficulty by grouping the queues upstream of a bottleneck and reducing their aggregate behavior to a simpler queue driven by exogenous arrivals. We adapt this idea to the node-server broadcast model.

For a relay node $v$, define the set of source-message streams that can feed the node bottleneck as
\begin{equation}
    \mathcal{I}_v
    =
    \left\{
    (s,m):
    s\in\mathcal{S}_m,\;
    v \text{ is reachable from } s
    \right\}.
\end{equation}
The aggregate exogenous input process associated with node $v$ is
\begin{equation}
    A^{\mathrm{br}}_v(t)
    =
    \sum_{(s,m)\in\mathcal{I}_v}
    A_s^m(t),
\end{equation}
and its mean arrival rate is
\begin{equation}
    \Lambda^{\mathrm{br}}_v
    =
    \sum_{(s,m)\in\mathcal{I}_v}
    \lambda_s^m .
\end{equation}
In a connected broadcast graph, every node is reachable from every source, and thus $\Lambda^{\mathrm{br}}_v$ becomes the aggregate coded-packet injection rate. This reflects the route-free forwarding rule: after an innovative first reception, a relay may enqueue and forward the coded packet even if it is not on the shortest path to a particular destination.

The bottleneck traffic intensity at node $v$ is
\begin{equation}
    \rho^{\mathrm{br}}_v
    =
    \frac{\Lambda^{\mathrm{br}}_v}{\mu_v}.
\end{equation}
The reduced bottleneck queue is stable only if
\begin{equation}
    \rho^{\mathrm{br}}_v < 1.
\end{equation}

\begin{theorem}[Node-bottleneck reduction]
\label{thm:node_bottleneck_reduction}
Consider a relay node $v$ and the aggregate input process $A^{\mathrm{br}}_v(t)$. Let $Q^{\mathrm{red}}_v(t)$ be a reduced single-server queue with service rate $\mu_v$ and evolution
\begin{equation}
    Q^{\mathrm{red}}_v(t+1)
    =
    \left(Q^{\mathrm{red}}_v(t)-\mu_v\right)^+
    +
    A^{\mathrm{br}}_v(t).
\end{equation}
Let $S_v(t)$ denote a relaxed aggregate upstream work process associated with
node $v$. This process counts all exogenous coded-packet arrivals from
source-message streams that can eventually require service from $v$, and it
relaxes the propagation and intermediate-queue constraints before packets reach
$v$. If
\begin{equation}
    Q^{\mathrm{red}}_v(0)=S_v(0)=0,
\end{equation}
then, for all $t\geq 0$,
\begin{equation}
    Q^{\mathrm{red}}_v(t)
    \leq
    S_v(t).
\end{equation}
\end{theorem}

\begin{proof}
The proof follows the sample-path reduction argument used for bottleneck reduction in fixed-route multi-hop wireless networks, adapted from a link bottleneck to a node-service bottleneck. Let $I_v(t)$ denote the number of packets served by node $v$ at slot $t$. Since node $v$ has service rate $\mu_v$,
\begin{equation}
    I_v(t)\leq \mu_v.
\end{equation}
The aggregate upstream work evolves as
\begin{equation}
    S_v(t+1)
    =
    S_v(t)-I_v(t)+A^{\mathrm{br}}_v(t).
\end{equation}
The actual system cannot serve more work from the aggregate than it contains, so
\begin{equation}
    S_v(t)\geq I_v(t).
\end{equation}

We prove the result by induction. The claim holds at $t=0$ because both systems are initially empty. Suppose
\begin{equation}
    Q^{\mathrm{red}}_v(t)\leq S_v(t).
\end{equation}
If $Q^{\mathrm{red}}_v(t)\geq\mu_v$, then
\begin{align}
    Q^{\mathrm{red}}_v(t+1)
    &=
    Q^{\mathrm{red}}_v(t)-\mu_v+A^{\mathrm{br}}_v(t) \\
    &\leq
    S_v(t)-\mu_v+A^{\mathrm{br}}_v(t) \\
    &\leq
    S_v(t)-I_v(t)+A^{\mathrm{br}}_v(t) \\
    &=
    S_v(t+1).
\end{align}
If $Q^{\mathrm{red}}_v(t)<\mu_v$, then
\begin{align}
    Q^{\mathrm{red}}_v(t+1)
    &=
    A^{\mathrm{br}}_v(t) \\
    &\leq
    S_v(t)-I_v(t)+A^{\mathrm{br}}_v(t) \\
    &=
    S_v(t+1),
\end{align}
where the inequality follows from $S_v(t)\geq I_v(t)$. Thus, in both cases,
\begin{equation}
    Q^{\mathrm{red}}_v(t+1)\leq S_v(t+1).
\end{equation}
By induction, the result holds for all $t\geq0$.
\end{proof}
The reduced queue is used to approximate the bottleneck pressure induced by
aggregate exogenous arrivals. It should not be interpreted as the exact queue
length at node $v$.

Theorem~\ref{thm:node_bottleneck_reduction} provides a sample-path reduction
that isolates the bottleneck pressure created by aggregate exogenous arrivals.
In the remainder of this paper, the corresponding reduced G/D/1 queue is used
as a tractable queueing-delay estimate rather than as an exact queue-length or
end-to-end delay bound.

\subsection{G/D/1 Bottleneck Delay Estimate}
\label{subsec:gd1_bottleneck_estimate}

The reduced queue at node $v$ is modeled as a G/D/1 queue. Let $c^2_{a,v}$ denote the squared coefficient of variation of the inter-arrival time process associated with $A^{\mathrm{br}}_v(t)$. Since the service time is deterministic, the squared coefficient of variation of the service time is zero. A Kingman-type approximation gives
\begin{equation}
    W^{\mathrm{G/D/1},\mathrm{br}}_{q,v}
    \approx
    \frac{c^2_{a,v}}{2}
    \frac{\rho^{\mathrm{br}}_v}{1-\rho^{\mathrm{br}}_v}
    \frac{1}{\mu_v},
    \quad
    \rho^{\mathrm{br}}_v<1.
\end{equation}
When $c^2_{a,v}=1$, the expression reduces to the M/D/1 waiting-time approximation
\begin{equation}
    W^{\mathrm{M/D/1},\mathrm{br}}_{q,v}
    =
    \frac{\rho^{\mathrm{br}}_v}
    {2\mu_v(1-\rho^{\mathrm{br}}_v)}
    =
    \frac{\Lambda^{\mathrm{br}}_v}
    {2\mu_v(\mu_v-\Lambda^{\mathrm{br}}_v)}.
\end{equation}

The bottleneck-reduction shortest-path delay estimate for coded packets generated by source $s$ for message $m$ and first received at destination $d$ is
\begin{equation}
    \widehat{D}^{\mathrm{br}}_{s,m,d}
    =
    \tau(R^{\mathrm{sp}}_{s,m,d})
    +
    \sum_{v\in
    \mathcal{V}_{\mathrm{tx}}
    (R^{\mathrm{sp}}_{s,m,d})}
    \left(
    \frac{1}{\mu_v}
    +
    W^{\mathrm{G/D/1},\mathrm{br}}_{q,v}
    \right).
\end{equation}
This expression is the main analytical estimate used in the numerical validation. The shortest path contributes the first-reception propagation component, while the bottleneck-reduction queue contributes the relay-node waiting component.

For a message-destination pair $(m,d)$, the source-weighted average estimate is
\begin{equation}
    \widehat{D}^{\mathrm{br}}_{m,d}
    =
    \sum_{s\in\mathcal{S}_m}
    \frac{\lambda_s^m}{\lambda_m}
    \widehat{D}^{\mathrm{br}}_{s,m,d},
    \quad
    \lambda_m>0.
\end{equation}

If one uncoded information unit is split into $G$ coded packets, then the equivalent-data-unit delay includes the time needed to generate enough coded chunks. 

A simple grouped-delay correction is
\begin{equation}
\widehat{D}^{\mathrm{br},G}_{m,d}
=
\widehat{D}^{\mathrm{br}}_{m,d}
+
\frac{G-1}{\lambda_m}.
\end{equation}
where the second term approximates the expected time from the first coded chunk in a group to the generation of the last chunk. When $G=1$, the estimate reduces to the packet-level delay estimate.

\subsection{Congestion-Cost Detour Approximation}
\label{subsec:congestion_cost_path}

\begin{figure}[!t]
\centering
\includegraphics[width=0.6\linewidth]{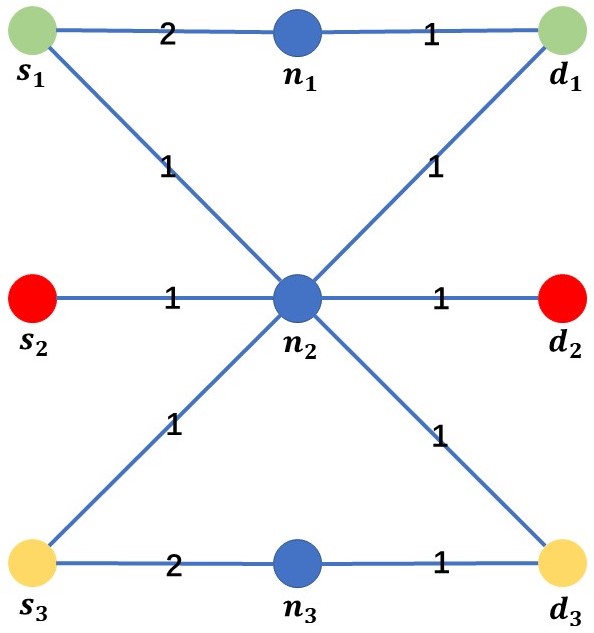}
\caption{9-node detour-enabled topology.}
\label{fig:topology_9}
\end{figure}

The bottleneck-reduction shortest-path estimate is the main analytical
estimate in this paper. However, under high load, the route-free broadcast
process may exhibit detour behavior: a packet may first reach a destination
through a longer but less congested relay sequence rather than through the
propagation-only shortest path. Fig.~\ref{fig:topology_9} illustrates this
motivation.

\begin{algorithm}[h]
\footnotesize
\DontPrintSemicolon

\textbf{Given:} $\mathbf{G}$, $\{\mathcal{S}_m,\mathcal{D}_m\}$,
$\{\lambda_s^m\}$, $\{\mu_v\}$, $\{\tau_e\}$,
$\beta$, $\eta$, $\epsilon$, $W_{\max}$, and $I_{\max}$.\\
Initialize $\mathcal{R}^{(0)}$ with propagation-only shortest routes,
$\widetilde W^{(0)}_{q,v}\leftarrow0$, and $\mathcal{H}\leftarrow\emptyset$.\\
\For{$i=0,1,\ldots,I_{\max}-1$}{
    Compute $\Lambda_v^{(i)}$, $\rho_v^{(i)}$, and the capped G/D/1
    penalties $\widehat W_{q,v}^{(i)}$.\\
    Apply damping to obtain $\widetilde W_{q,v}^{(i)}$.\\
    Update every equivalent route using the congestion-aware cost in
    the preceding equation.\\
    \If{$\mathcal{R}^{(i+1)}=\mathcal{R}^{(i)}$}{
        Return $\mathcal{R}^{(i+1)}$.\\
    }
    \If{$\mathcal{R}^{(i+1)}\in\mathcal{H}$}{
        Return the visited route set with minimum total estimated delay.\\
    }
    Add $\mathcal{R}^{(i+1)}$ to $\mathcal{H}$.\\
}
Return the visited route set with minimum total estimated delay.\\

\caption{\small Damped congestion-cost equivalent-route approximation.}
\label{alg:congestion_cost_path}
\end{algorithm}

In Fig.~\ref{fig:topology_9}, the propagation-shortest routes from the three
sources to their destinations all pass through the relay $n_2$. A fixed
shortest-path approximation therefore concentrates all three source streams on
one single-server relay, which may become unstable as the aggregate arrival
rate increases. In contrast, route-free broadcasting is not restricted to this
route set: packets from the upper and lower pairs may first arrive through the
side relays $n_1$ and $n_3$. These side paths have larger propagation delay but
avoid the congested central relay. This motivates a congestion-aware
equivalent-route approximation.

Let $\mathcal{R}^{(i)}=\{R^{(i)}_{s,m,d}:m\in\mathcal{M},\ s\in\mathcal{S}_m,\ d\in\mathcal{D}_m\}$ denote the equivalent route set at iteration $i$. The load induced at node $v$ is
\begin{equation}
    \Lambda^{(i)}_v
    =
    \sum_{m\in\mathcal{M}}
    \sum_{s\in\mathcal{S}_m}
    \lambda_s^m
    \mathbf{1}_{\left\{
    v\in
    \bigcup_{d\in\mathcal{D}_m}
    \mathcal{V}_{\mathrm{tx}}(R^{(i)}_{s,m,d})
    \right\}},
    \qquad
    \rho^{(i)}_v=\frac{\Lambda^{(i)}_v}{\mu_v}.
\end{equation}
Thus, a source-message stream is counted once at node $v$ if $v$ appears on at
least one current equivalent route from that source to a destination of the
same message.

The queueing penalty is capped near instability:
\begin{equation}
    \widehat{W}^{(i)}_{q,v}
    =
    \begin{cases}
    W^{(i)}_{q,v}, & \rho^{(i)}_v<1-\epsilon,\\
    W_{\max}, & \rho^{(i)}_v\geq 1-\epsilon,
    \end{cases}
\end{equation}
where $W^{(i)}_{q,v}$ is computed using the M/D/1 or G/D/1 approximation,
$\epsilon>0$ is a stability margin, and $W_{\max}$ avoids the singularity near
$\rho_v=1$. To reduce route oscillation, we use the smoothed penalty
\begin{equation}
    \widetilde{W}^{(i)}_{q,v}
    =
    (1-\eta)\widetilde{W}^{(i-1)}_{q,v}
    +
    \eta\widehat{W}^{(i)}_{q,v},
    \quad
    0<\eta\leq1.
\end{equation}

Given the smoothed penalties, each equivalent route is updated by minimizing
the congestion-aware route cost:
\begin{equation}
    R^{(i+1)}_{s,m,d}
    =
    \arg\min_{R\in\mathcal{R}_{sd}}
    \left[
    \tau(R)
    +
    \beta
    \sum_{v\in\mathcal{V}_{\mathrm{tx}}(R)}
    \widetilde{W}^{(i)}_{q,v}
    \right],
\end{equation}
where $\beta\geq0$ controls the weight of the queueing penalty. Ties are
broken deterministically. The iteration stops when the route set reaches a
fixed point, repeats a previously visited route set, or reaches the maximum
iteration number. If a cycle occurs, the visited route set with the smallest
total estimated delay is returned. This procedure is a theoretical
route-update approximation, not a globally convergent routing algorithm.

\section{Simulation Evaluation}\label{sec:num}

This section validates the bottleneck-reduction estimate on random
topologies and the congestion-cost detour approximation on structured
topologies.

\begin{table}[htbp]
    \centering
    \caption{IEEE 802.11bd parameters used for time calibration}
    \label{tab:80211bd_parameters}
    \resizebox{\columnwidth}{!}{
    \begin{tabular}{lll}
        \hline
        \textbf{Parameter} & \textbf{Symbol} & \textbf{Value} \\
        \hline
        Carrier frequency
            & $f_c$
            & $5.9$ GHz \\

        Channel bandwidth
            & $W$
            & $10$ MHz \\

        Packet size
            & $L$
            & $350$ bytes \\

        Modulation and coding scheme
            & MCS
            & MCS 2, QPSK, coding rate $1/2$ \\

        Coded data bits per OFDM symbol
            & $N_{\mathrm{DBPS}}$
            & $48$ bits \\

        OFDM symbol duration
            & $T_{\mathrm{sym}}$
            & $8\,\mu\mathrm{s}$ \\

        Preamble and SIGNAL duration
            & $T_{\mathrm{PHY}}$
            & $40\,\mu\mathrm{s}$ \\

        Arbitration inter-frame space
            & $T_{\mathrm{AIFS}}$
            & $110\,\mu\mathrm{s}$ \\

        MAC backoff mini-slot
            & $\sigma$
            & $13\,\mu\mathrm{s}$ \\

        Maximum contention window
            & $CW$
            & $15$ \\

        Initial backoff counter
            & $B$
            & $U\{0,1,\ldots,15\}$ \\
        \hline
    \end{tabular}}
\end{table}

\begin{table}[htbp]
\centering
\caption{Random-topology settings used in the main validation.}
\label{tab:random_topology_stats}
\scriptsize
\resizebox{\columnwidth}{!}{
\begin{tabular}{c|c|c|c|c|c|c}
\hline
\textbf{Topology}
& \textbf{Nodes}
& \textbf{Src./Dst.}
& \textbf{Range (m)}
& \textbf{Edges}
& \textbf{Density}
& \textbf{Seed} \\
\hline
30 sparse
& 30
& 3/3
& 250
& 63
& 0.145
& 1025 \\

50 medium
& 50
& 3/3
& 230
& 182
& 0.149
& 1979 \\

100 dense
& 100
& 10/10
& 205
& 517
& 0.104
& 3000 \\

150 dense
& 150
& 10/10
& 190
& 1028
& 0.092
& 4000 \\
\hline
\end{tabular}}
\end{table}

\begin{table}[htbp]
\centering
\caption{Structured detour-validation settings.}
\label{tab:detour_topology_settings}
\scriptsize
\resizebox{\columnwidth}{!}{
\begin{tabular}{c|c|c|c|c|c|c|c}
\hline
\textbf{Topology}
& \textbf{Nodes}
& \textbf{Src./Dst.}
& \textbf{Directed edges}
& \textbf{Range}
& \textbf{Seed}
& \textbf{Sim./warm-up}
& \textbf{Reps.} \\
\hline
9-node
& 9
& 3/3
& 10
& Structured
& 
& 5000/800
& 10 \\

50-node
& 50
& 6/6
& 58
& 230 m
& 3018
& 5000/800
& 10 \\
\hline
\end{tabular}}
\end{table}

\subsection{Simulation Setup}

All experiments use the graph-level queueing model in
Section~\ref{sec:sys}. Each node has one FIFO outgoing queue
and serves at most one coded packet per simulation slot. The
slot duration is calibrated using the IEEE 802.11bd simulation
profile in~\cite{wu2023adaptive}, as summarized in
Table~\ref{tab:80211bd_parameters}. Packet repetitions are
disabled because the analytical model considers one
transmission opportunity for each served coded packet.

The Physical Layer Convergence Procedure Protocol Data Unit(PPDU) duration is
\begin{equation}
    T_{\mathrm{PPDU}}
    =
    40+
    8\left\lceil
    \frac{16+8\times350+6}{48}
    \right\rceil
    =
    512\,\mu\mathrm{s}.
\end{equation}
Since $\mathbb{E}[B]=7.5$, one service slot is
\begin{equation}
    \Delta_{\mathrm{slot}}
    =
    110+7.5\times13+512
    =
    719.5\,\mu\mathrm{s}
    \approx0.72\,\mathrm{ms}
\end{equation}
and $\mu_v=1$ packet/slot.

To represent heterogeneous baseline link conditions, each
directed link is assigned

\begin{equation}
    X_e
    \sim
    \operatorname{Lognormal}
    \left(\ln 3,\,0.55^2\right),
\end{equation}
and
\begin{equation}
    \tau_e
    =
    \min\left\{
        10,
        \max\left\{
            1,
            \left\lceil X_e\right\rceil
        \right\}
    \right\}.
    \label{eq:link_delay_sampling}
\end{equation}
Thus,
\begin{equation}
    T_e^{\mathrm{link}}
    =
    \tau_e\Delta_{\mathrm{slot}}
    \in
    [0.72,\,7.20]\ \mathrm{ms}.
    \label{eq:physical_link_delay_range}
\end{equation}
The sampled link delay remains fixed, while load-dependent waiting is
generated separately by the FIFO queues. All reported delays are
converted to milliseconds.

The main validation uses four connected random geometric graphs in a
$1299\,\mathrm{m}\times750\,\mathrm{m}$ region. Two nodes are connected
when their distance does not exceed the scenario-specific communication
range.

\begin{figure}[t]
\centering
\includegraphics[width=0.95\linewidth]
{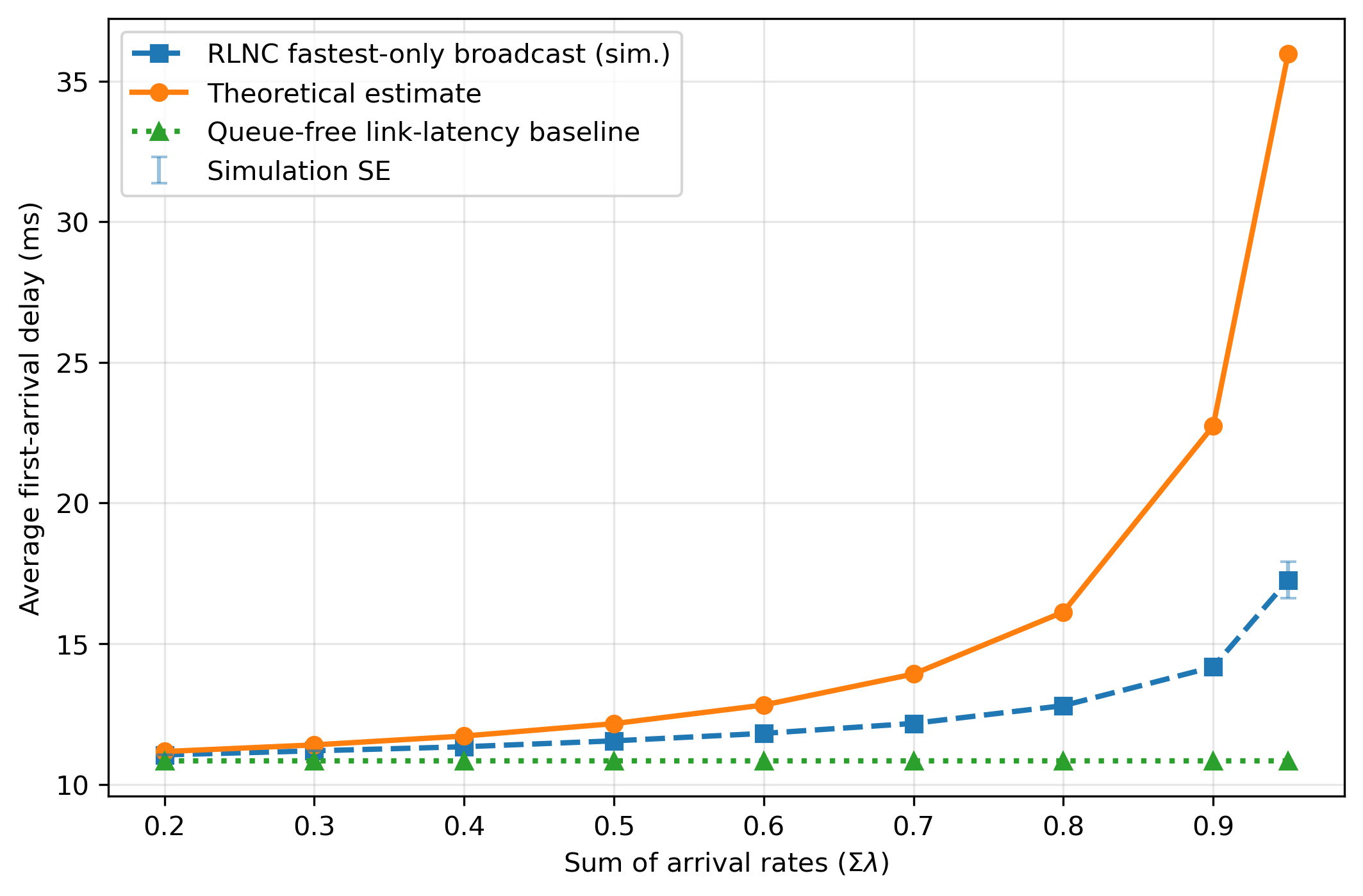}
\caption{Avg. delay on the 30-node sparse topology.}
\label{fig:general_30}
\end{figure}

\begin{figure}[t]
\centering
\includegraphics[width=0.95\linewidth]
{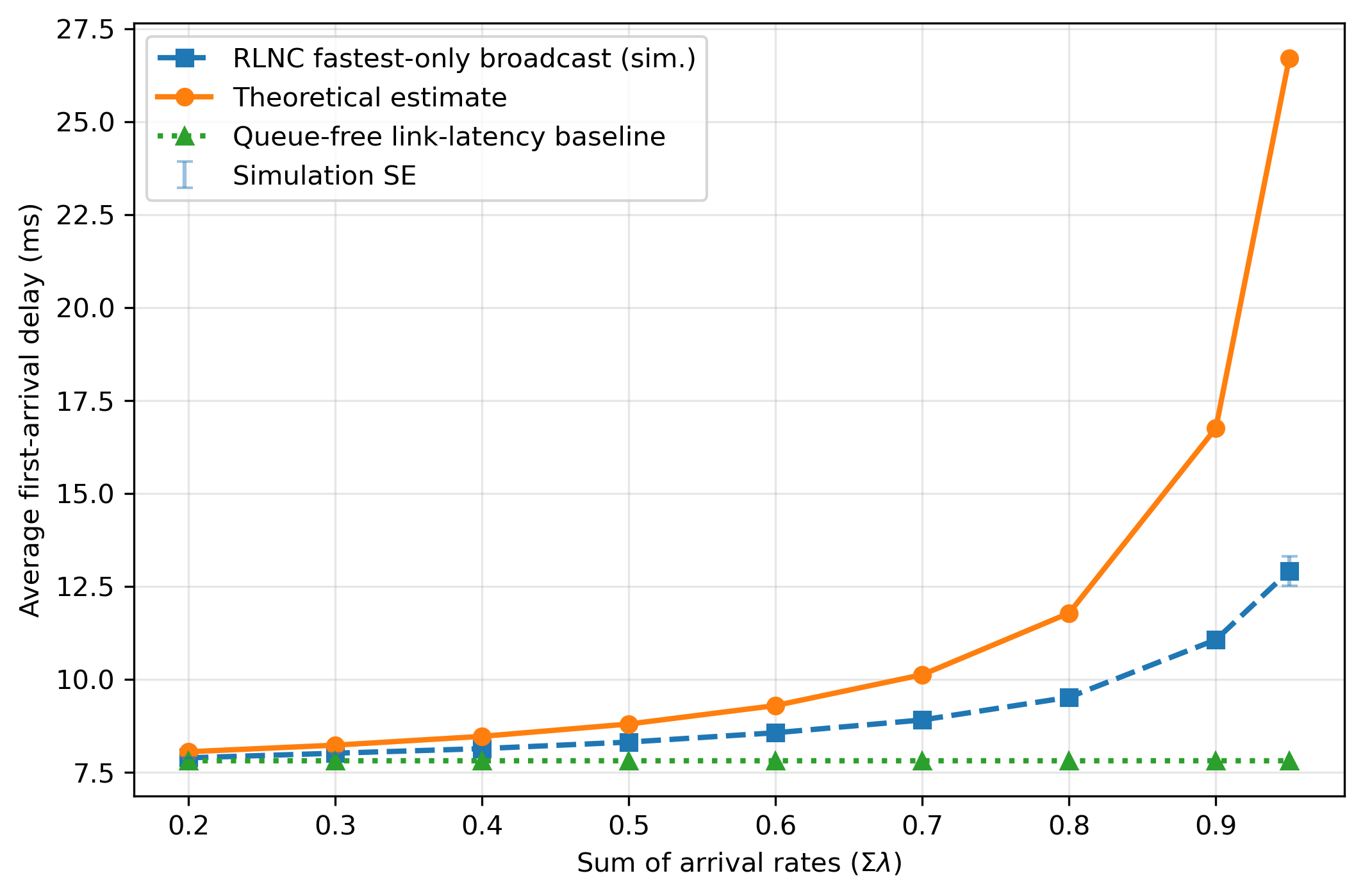}
\caption{Avg. delay on the 50-node medium-density topology.}
\label{fig:general_50}
\end{figure}

\begin{figure}[t]
\centering
\includegraphics[width=0.95\linewidth]
{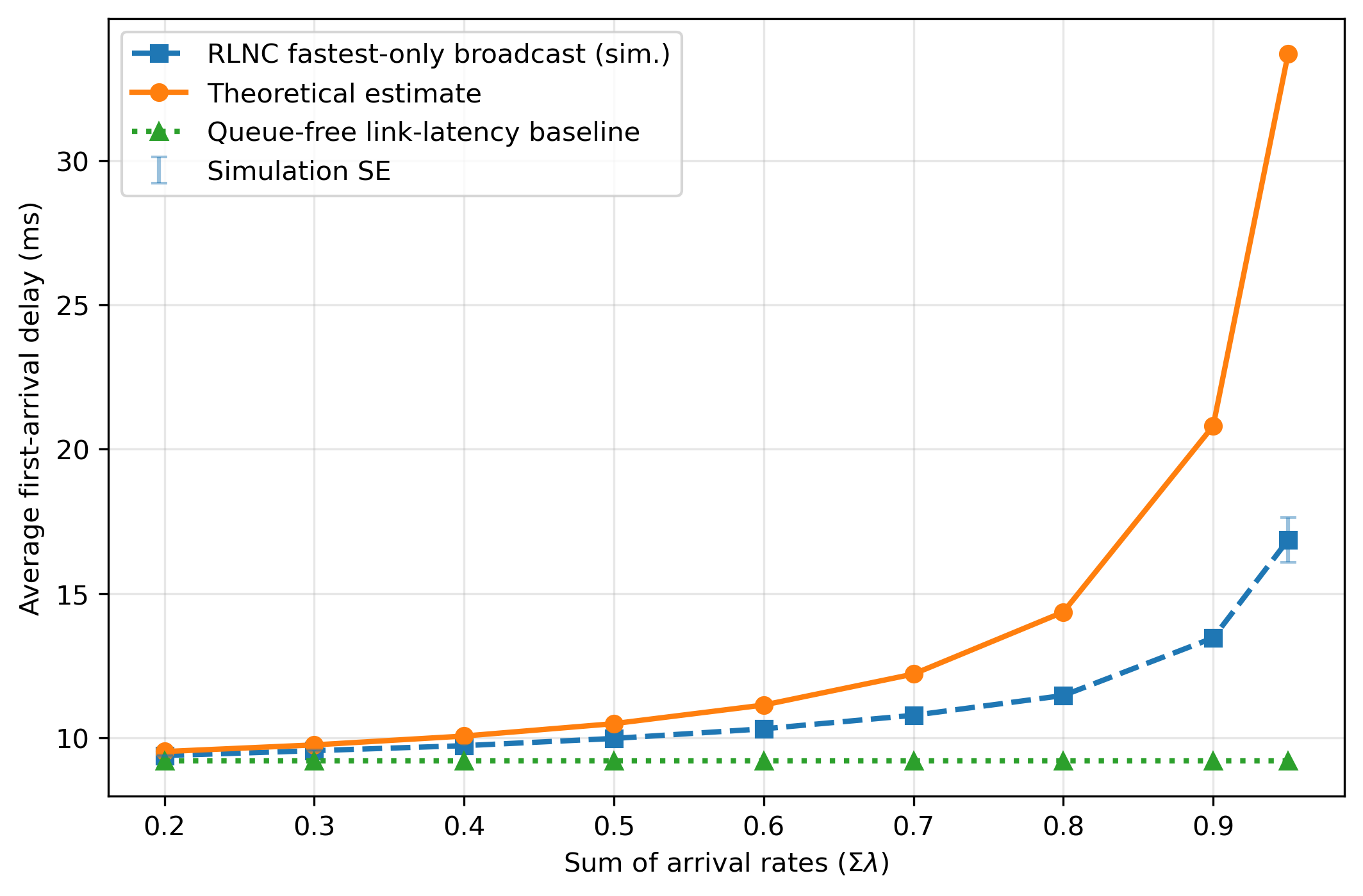}
\caption{Avg. delay on the 100-node dense topology.}
\label{fig:general_100}
\end{figure}

\begin{figure}[t]
\centering
\includegraphics[width=0.95\linewidth]
{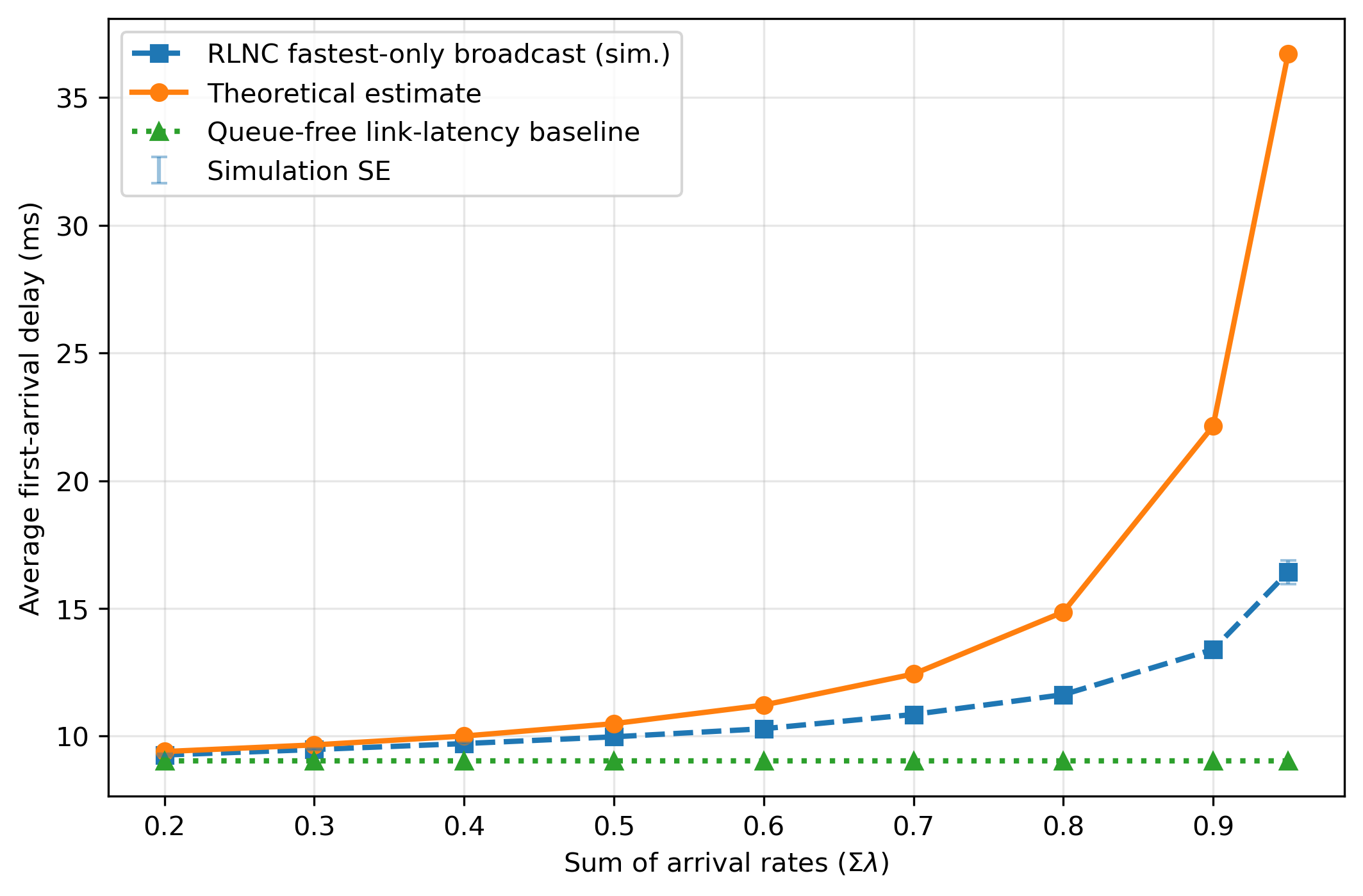}
\caption{Avg. delay on the 150-node dense topology.}
\label{fig:general_150}
\end{figure}

Sources and destinations are disjoint, and every source broadcasts to
every destination. Source arrivals are independent Bernoulli processes,
with
\begin{equation}
    \lambda_s
    =
    \frac{w_s\Sigma\lambda}{\sum_{j\in\mathcal{S}}w_j},
    \qquad
    w_s\sim U\{100,\ldots,149\},
\end{equation}
using splitting seed 1234. The aggregate rate is
\begin{equation}
    \Sigma\lambda
    \in
    \{0.20,0.30,0.40,0.50,0.60,0.70,0.80,0.90,0.95\}.
\end{equation}
Each operating point uses 5000 slots, a 600-slot warm-up, and 10
independent repetitions. Error bars show simulation standard errors.

The detour validation uses the structured 9-node topology in
Fig.~\ref{fig:topology_9} and a matched-disjoint 50-node topology.

In the 9-node case, all sources use
\begin{equation}
    \lambda_i=\frac{\Sigma\lambda}{3},
\end{equation}
and delay is averaged over the two detour-capable pairs
$(s_1,d_1)$ and $(s_3,d_3)$. The 50-node case contains six paired
flows, one central bottleneck, and three side corridors shared by two
flows each, with
\begin{equation}
    \lambda_i=\frac{\Sigma\lambda}{6}.
\end{equation}
All six flows are included in the reported delay. For both cases,
\begin{equation}
    \Sigma\lambda
    \in
    \{0.20,0.30,\ldots,2.50\}.
\end{equation}

The detour analysis uses
\begin{equation}
    \beta=1,
    \qquad
    c_a^2=0.4,
    \qquad
    \mu_v=1,
\end{equation}
and reports
\begin{equation}
    \widehat{D}_{\mathrm{final}}
    =
    \min\left\{
        \widehat{D}_{\mathrm{general}},
        \widehat{D}_{\mathrm{detour}}
    \right\}.
\end{equation}

\subsection{Random-Topology Validation}
\label{subsec:random_validation}

Figs.~\ref{fig:general_30}--\ref{fig:general_150} compare the
route-free simulation with the bottleneck-reduction estimate and the
queue-free baseline. The baseline is load-independent, whereas both
queueing curves increase with $\Sigma\lambda$. The estimate follows the
simulation at low and moderate load but becomes conservative near
$\Sigma\lambda=1$ because the G/D/1 penalty grows as
$\rho_v/(1-\rho_v)$.

The simulated delay ranges are approximately $11.05$--$17.26$ ms,
$7.89$--$12.91$ ms, $9.39$--$16.86$ ms, and $9.26$--$16.42$ ms for
the 30, 50, 100, and 150 node cases, respectively. Across these
cases, the RMSE is $5.060$--$7.470$ ms, the normalized RMSE is
$0.421$--$0.491$, and the mean symmetric relative error is
$0.182$--$0.204$. The differences mainly reflect connectivity,
source--destination placement, and weighted path lengths rather than
node count alone.

\subsection{Topology with Detour Validation}
\label{subsec:detour_validation}

\begin{figure}[t]
\centering
\includegraphics[width=0.96\linewidth]
{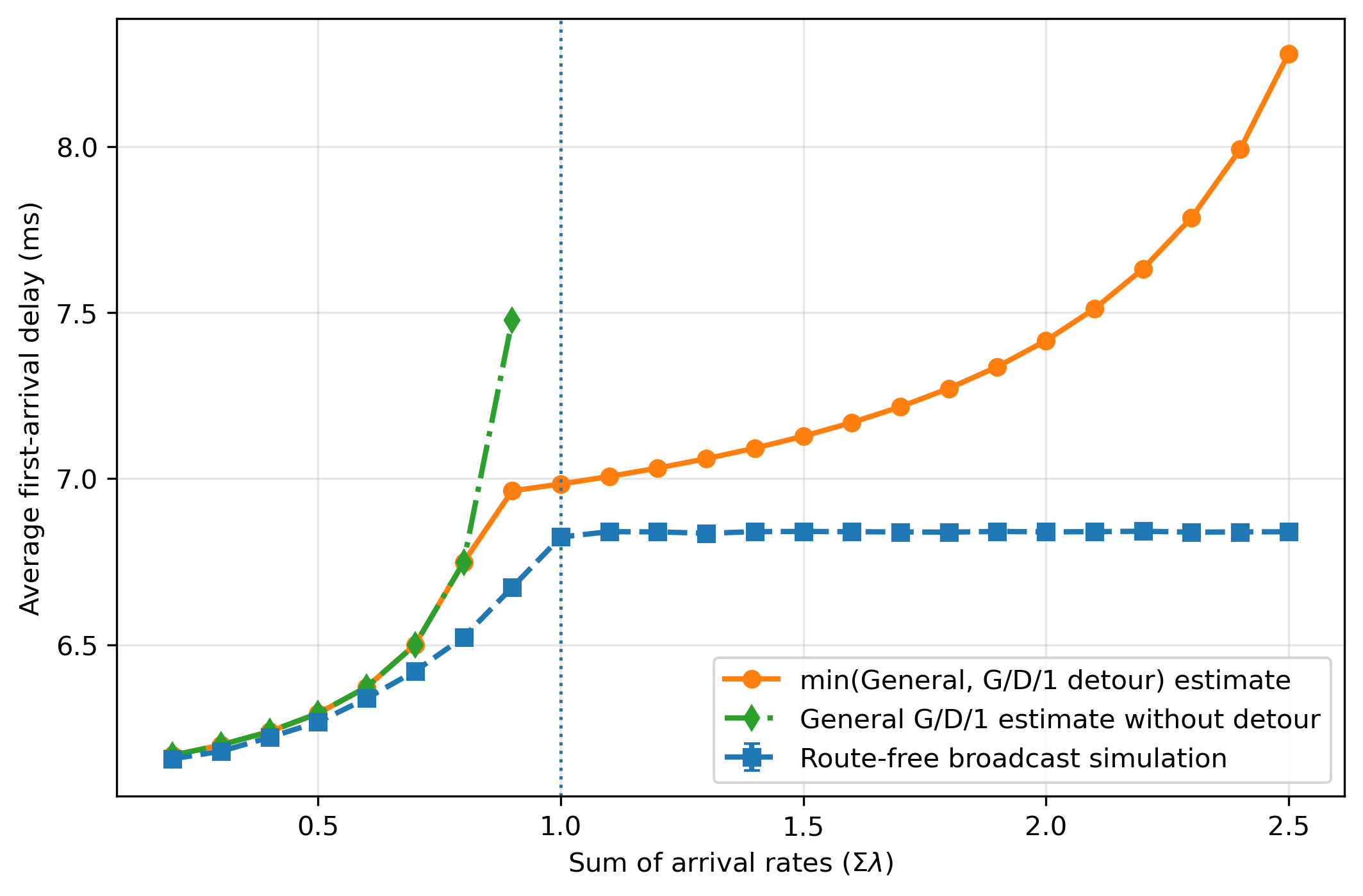}
\caption{Avg. delay on the 9-node topology with detour.}
\label{fig:detour_9_new}
\end{figure}

\begin{figure}[t]
\centering
\includegraphics[width=\linewidth]
{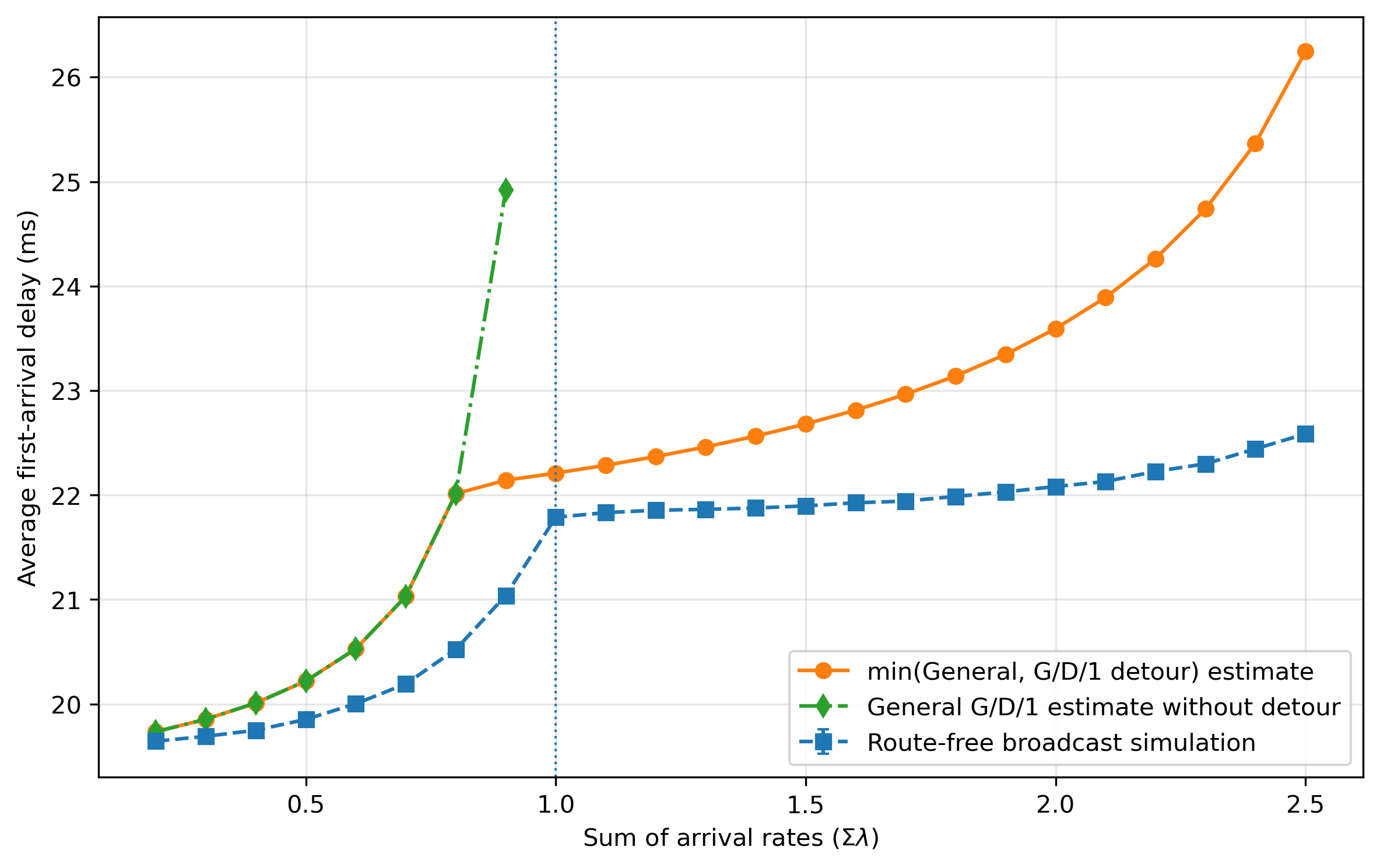}
\caption{Avg. delay on the 50-node topology with detour.}
\label{fig:detour_50_new}
\end{figure}

Figs.~\ref{fig:detour_9_new}--\ref{fig:detour_50_new} compare the
route-free simulation with the fixed shortest-route G/D/1 estimate
and $\widehat D_{\mathrm{final}}$.

In the 9-node case, the no-detour estimate becomes unstable near
$\Sigma\lambda=1$, whereas the simulation remains close to $6.84$ ms
after switching to side paths. Each observed side relay receives only
one Bernoulli source with
\begin{equation}
    \lambda_i
    =
    \frac{\Sigma\lambda}{3}
    \leq
    \frac{2.5}{3}
    <1.
\end{equation}
Since one source generates at most one packet per slot and the relay
serves one packet per slot, no persistent side-path queue forms. The
middle flow contributes background congestion but is excluded from the
reported delay.

In the 50-node case, each side corridor aggregates two independent
sources. Its queueing delay therefore continues to grow after detouring.
The simulated delay increases from about $19.64$ to $22.55$ ms, while
the final estimate increases from about $19.73$ to $26.25$ ms.

\subsection{Discussion}
\label{subsec:simulation_discussion}

The bottleneck-reduction approximation captures the low- and
moderate-load trend of the general cases but becomes conservative near
the stability boundary. The structured cases further show that longer
paths can avoid an unstable central bottleneck.

The remaining gap arises because the detour theory is a
single-equivalent-route G/D/1 approximation. It does not explicitly represent
the fraction of packets using each detour path, path-dependent arrival
variability, or correlations among parallel paths and successive relay
queues. It also replaces the minimum over several correlated
first-arrival times with one representative-route cost.

Despite these simplifications, the approximation identifies the
shortest-route instability, captures the propagation--congestion
tradeoff, and produces a finite detour estimate after the fixed-route
model loses stability.

\section{Conclusion}\label{sec:con}

In this paper, we studied packet delay in M-to-N ad-hoc broadcast networks enabled by random linear network coding. We proved that, under propagation-only dissemination, the fastest-only broadcast rule yields the same first-reception time as the multi-source shortest-path distance, which provides a rigorous lower bound on actual end-to-end delay. We then developed a bottleneck-aware queueing approximation based on fixed-route M/D/1 and G/D/1 models. For some topologies where propagation-only shortest paths concentrate on a congested relay, we further introduced a congestion-cost path approximation to capture congestion-induced route diversity.

\bibliographystyle{IEEEtran}
\bibliography{Ref}

@ARTICLE{1705002,
  author={Ho, T. and Medard, M. and Koetter, R. and Karger, D.R. and Effros, M. and Shi, J. and Leong, B.},
  journal={IEEE Transactions on Information Theory}, 
  title={A Random Linear Network Coding Approach to Multicast}, 
  year={2006},
  volume={52},
  number={10},
  pages={4413-4430},
  doi={10.1109/TIT.2006.881746}}

@INPROCEEDINGS{1228459,
  author={Ho, T. and Koetter, R. and Medard, M. and Karger, D.R. and Effros, M.},
  booktitle={IEEE International Symposium on Information Theory, 2003. Proceedings.}, 
  title={The benefits of coding over routing in a randomized setting}, 
  year={2003},
  volume={},
  number={},
  pages={442-},
  doi={10.1109/ISIT.2003.1228459}}

@INPROCEEDINGS{4753100,
  author={Fujimura, Atsushi and Oh, Soon Y. and Gerla, Mario},
  booktitle={MILCOM 2008 - 2008 IEEE Military Communications Conference}, 
  title={Network coding vs. erasure coding: Reliable multicast in ad hoc networks}, 
  year={2008},
  volume={},
  number={},
  pages={1-7},
  doi={10.1109/MILCOM.2008.4753100}}

@INPROCEEDINGS{4146698,
  author={Fragouli, C. and Widmer, J. and Le Boudec, J.-Y.},
  booktitle={Proceedings IEEE INFOCOM 2006. 25TH IEEE International Conference on Computer Communications}, 
  title={A Network Coding Approach to Energy Efficient Broadcasting: From Theory to Practice}, 
  year={2006},
  volume={},
  number={},
  pages={1-11},
  doi={10.1109/INFOCOM.2006.45}}

@ARTICLE{9201370,
  author={Song, Hao and Liu, Lingjia and Pudlewski, Scott M. and Bentley, Elizabeth Serena},
  journal={IEEE Transactions on Wireless Communications}, 
  title={Random Network Coding Enabled Routing Protocol in Unmanned Aerial Vehicle Networks}, 
  year={2020},
  volume={19},
  number={12},
  pages={8382-8395},
  doi={10.1109/TWC.2020.3022399}}

@ARTICLE{5688207,
  author={Gupta, Gagan Raj and Shroff, Ness B.},
  journal={IEEE/ACM Transactions on Networking}, 
  title={Delay Analysis and Optimality of Scheduling Policies for Multihop Wireless Networks}, 
  year={2011},
  volume={19},
  number={1},
  pages={129-141},
  doi={10.1109/TNET.2010.2095506}}

@ARTICLE{8408468,
  author={Ye, Qiang and Zhuang, Weihua and Li, Xu and Rao, Jaya},
  journal={IEEE Internet of Things Journal}, 
  title={End-to-End Delay Modeling for Embedded VNF Chains in 5G Core Networks}, 
  year={2019},
  volume={6},
  number={1},
  pages={692-704},
  doi={10.1109/JIOT.2018.2853708}}

@article{10.1145/1143549.1143704,
title = {Queuing network models for delay analysis of multihop wireless ad hoc networks},
journal = {Ad Hoc Networks},
volume = {7},
number = {1},
pages = {79-97},
year = {2009},
issn = {1570-8705},
doi = {https://doi.org/10.1016/j.adhoc.2007.12.001},
author = {Nabhendra Bisnik and Alhussein A. Abouzeid},
}

@INPROCEEDINGS{5205612,
  author={Drinea, Eleni and Fragouli, Christina and Keller, Lorenzo},
  booktitle={2009 IEEE International Symposium on Information Theory}, 
  title={Delay with network coding and feedback}, 
  year={2009},
  volume={},
  number={},
  pages={844-848},
  doi={10.1109/ISIT.2009.5205612}}

@ARTICLE{850663,
  author={Ahlswede, R. and Ning Cai and Li, S.-Y.R. and Yeung, R.W.},
  journal={IEEE Transactions on Information Theory}, 
  title={Network information flow}, 
  year={2000},
  volume={46},
  number={4},
  pages={1204-1216},
  doi={10.1109/18.850663}}

@book{1941130,
author = {Neely, Michael J.},
title = {Stochastic Network Optimization with Application to Communication and Queueing Systems},
year = {2010},
isbn = {160845455X},
publisher = {Morgan and Claypool Publishers}
}

@INPROCEEDINGS{4215785,
  author={Li, L. and Ramjee, R. and Buddhikot, M. and Miller, S.},
  booktitle={IEEE INFOCOM 2007 - 26th IEEE International Conference on Computer Communications}, 
  title={Network Coding-Based Broadcast in Mobile Ad-hoc Networks}, 
  year={2007},
  volume={},
  number={},
  pages={1739-1747},
  doi={10.1109/INFCOM.2007.203}}

@inproceedings{10.1145/1282380.1282400,
author = {Chachulski, Szymon and Jennings, Michael and Katti, Sachin and Katabi, Dina},
title = {Trading structure for randomness in wireless opportunistic routing},
year = {2007},
isbn = {9781595937131},
publisher = {Association for Computing Machinery},
address = {New York, NY, USA},
doi = {10.1145/1282380.1282400},
booktitle = {Proceedings of the 2007 Conference on Applications, Technologies, Architectures, and Protocols for Computer Communications},
pages = {169–180},
numpages = {12},
location = {Kyoto, Japan},
series = {SIGCOMM '07}
}

@ARTICLE{1435642,
  author={Neely, M.J. and Modiano, E.},
  journal={IEEE Transactions on Information Theory}, 
  title={Capacity and delay tradeoffs for ad hoc mobile networks}, 
  year={2005},
  volume={51},
  number={6},
  pages={1917-1937},
  doi={10.1109/TIT.2005.847717}}

@article{Kingman_1961, 
title={The single server queue in heavy traffic}, 
volume={57}, 
DOI={10.1017/S0305004100036094}, 
number={4},
journal={Mathematical Proceedings of the Cambridge Philosophical Society}, 
author={Kingman, J. F. C.}, 
year={1961}, 
pages={902–904}
}

@INPROCEEDINGS{6363697,
  author={Kunz, Thomas and Mahmood, Kashif and Li, Li},
  booktitle={2012 IEEE International Conference on Communications (ICC)}, 
  title={Broadcasting in multihop wireless networks: The case for multi-source network coding}, 
  year={2012},
  volume={},
  number={},
  pages={5157-5162},
  doi={10.1109/ICC.2012.6363697}}

@inproceedings{10.1145/3274783.3274849,
author = {Herrmann, Carsten and Mager, Fabian and Zimmerling, Marco},
title = {Mixer: Efficient Many-to-All Broadcast in Dynamic Wireless Mesh Networks},
year = {2018},
isbn = {9781450359528},
publisher = {Association for Computing Machinery},
address = {New York, NY, USA},
doi = {10.1145/3274783.3274849},
booktitle = {Proceedings of the 16th ACM Conference on Embedded Networked Sensor Systems},
pages = {145–158},
numpages = {14},
location = {Shenzhen, China},
series = {SenSys '18}
}

@ARTICLE{5403547,
  author={Asterjadhi, Alfred and Fasolo, Elena and Rossi, Michele and Widmer, Joerg and Zorzi, Michele},
  journal={IEEE Transactions on Wireless Communications}, 
  title={Toward network coding-based protocols for data broadcasting in wireless Ad Hoc networks}, 
  year={2010},
  volume={9},
  number={2},
  pages={662-673},
  doi={10.1109/TWC.2010.02.081057}}

@inproceedings{10.1145/1280940.1280959,
author = {Rahnavard, Nazanin and Vellambi, Badri N. and Fekri, Faramarz},
title = {Efficient broadcasting via rateless coding in multihop wireless networks with local information},
year = {2007},
isbn = {9781595936950},
publisher = {Association for Computing Machinery},
address = {New York, NY, USA},
doi = {10.1145/1280940.1280959},
booktitle = {Proceedings of the 2007 International Conference on Wireless Communications and Mobile Computing},
pages = {85–90},
numpages = {6},
location = {Honolulu, Hawaii, USA},
series = {IWCMC '07}
}

@INPROCEEDINGS{5426175,
  author={Yang, Zhenyu and Li, Ming and Lou, Wenjing},
  booktitle={GLOBECOM 2009 - 2009 IEEE Global Telecommunications Conference}, 
  title={R-Code: Network Coding Based Reliable Broadcast in Wireless Mesh Networks with Unreliable Links}, 
  year={2009},
  volume={},
  number={},
  pages={1-6},
  doi={10.1109/GLOCOM.2009.5426175}}

@misc{10.1145/972374.972387,
title={Opportunistic routing in multi-hop wireless networks}, 
url={http://hdl.handle.net/1721.1/34115}, 
author={Biswas, Sanjit Zubin}, 
year={2005} }

@INPROCEEDINGS{5061904,
  author={Laufer, R. and Dubois-Ferriere, H. and Kleinrock, L.},
  booktitle={IEEE INFOCOM 2009}, 
  title={Multirate Anypath Routing in Wireless Mesh Networks}, 
  year={2009},
  volume={},
  number={},
  pages={37-45},
  doi={10.1109/INFCOM.2009.5061904}}

@ARTICLE{6553237,
  author={Swapna, B. T. and Eryilmaz, Atilla and Shroff, Ness B.},
  journal={IEEE Transactions on Information Theory}, 
  title={Throughput-Delay Analysis of Random Linear Network Coding for Wireless Broadcasting}, 
  year={2013},
  volume={59},
  number={10},
  pages={6328-6341},
  doi={10.1109/TIT.2013.2271895}}

@ARTICLE{7858628,
  author={Chatzigeorgiou, Ioannis and Tassi, Andrea},
  journal={IEEE Transactions on Vehicular Technology}, 
  title={Decoding Delay Performance of Random Linear Network Coding for Broadcast}, 
  year={2017},
  volume={66},
  number={8},
  pages={7050-7060},
  doi={10.1109/TVT.2017.2670178}}

@ARTICLE{9112345,
  author={Su, Rina and Sun, Qifu Tyler and Zhang, Zhongshan},
  journal={IEEE Transactions on Communications}, 
  title={Delay-Complexity Trade-Off of Random Linear Network Coding in Wireless Broadcast}, 
  year={2020},
  volume={68},
  number={9},
  pages={5606-5618},
  doi={10.1109/TCOMM.2020.3001133}}

@ARTICLE{9944003,
  author={Su, Rina and Sun, Qifu Tyler and Zhang, Zhongshan and Li, Zongpeng},
  journal={IEEE Transactions on Communications}, 
  title={Completion Delay of Random Linear Network Coding in Full-Duplex Relay Networks}, 
  year={2022},
  volume={70},
  number={12},
  pages={7843-7857},
  doi={10.1109/TCOMM.2022.3220867}}

@article{PARANJOTHI2020102277,
title = {A survey on congestion detection and control in connected vehicles},
journal = {Ad Hoc Networks},
volume = {108},
pages = {102277},
year = {2020},
issn = {1570-8705},
doi = {10.1016/j.adhoc.2020.102277},
author = {Anirudh Paranjothi and Mohammad S. Khan and Sherali Zeadally}
}

@article{SHAHWANI2022100420,
title = {A comprehensive survey on data dissemination in Vehicular Ad Hoc Networks},
journal = {Vehicular Communications},
volume = {34},
pages = {100420},
year = {2022},
issn = {2214-2096},
doi = {10.1016/j.vehcom.2021.100420},
author = {Hamayoun Shahwani and Syed {Attique Shah} and Muhammad Ashraf and Muhammad Akram and Jaehoon (Paul) Jeong and Jitae Shin},
}

@ARTICLE{9775109,
  author={Pei, Zhonghui and Chen, Wei and Li, Changzhen and Du, Luyao and Liu, Huiheng and Wang, Xiaojun},
  journal={IEEE Sensors Journal}, 
  title={Analysis and Optimization of Multihop Broadcast Communication in the Internet of Vehicles Based on C-V2X Mode 4}, 
  year={2022},
  volume={22},
  number={12},
  pages={12428-12443},
  doi={10.1109/JSEN.2022.3175158}}

@article{wu2023adaptive,
  author  = {Zhuofei Wu and Stefania Bartoletti and
             Vincent Martinez and Alessandro Bazzi},
  title   = {Adaptive Repetitions Strategies in
             {IEEE} 802.11bd},
  journal = {IEEE Transactions on Vehicular Technology},
  year    = {2023},
  doi     = {10.1109/TVT.2023.3241865}
}
\end{document}